\RequirePackage{fixltx2e}

\documentclass[aip,apl, floatfix,reprint,groupedaddress,showpacs,showkeys,amsmath,amssymb,]{revtex4-1}

\usepackage{amsfonts}%
\usepackage[ngerman,english]{babel}
\usepackage[utf8]{inputenc}
\usepackage{graphicx}
\usepackage{dcolumn}
\usepackage{bm}
\usepackage{color}

\begin{document}


\title{Fundamental absorption edges in heteroepitaxial YBiO\textsubscript{3} thin films}


\author{Marcus Jenderka}
	\email[Corresponding Author: ]{marcus.jenderka@physik.uni-leipzig.de}
\author{Steffen Richter}
\author{Michael Lorenz}
\author{Marius Grundmann}
	
\homepage[]{http://research.uni-leipzig.de/hlp/}
\affiliation{%
Institut für Experimentelle Physik II, Universität Leipzig\\
Linnéstraße 5, D-04103 Leipzig (Germany)
}%


\date{\today}

\begin{abstract}
The dielectric function of heteroepitaxial YBiO$_3$ grown on $a$-Al$_2$O$_3$ single crystals via pulsed laser deposition is determined in the spectral range from 0.03 eV to 4.5 eV by simultaneous modeling of spectroscopic ellipsometry and optical transmission data of YBiO$_3$ films of different thickness. The (111)-oriented YBiO$_3$ films are nominally unstrained and crystallize in a defective fluorite-type structure with $Fm\bar{3}m$ space group. From the calculated absorption spectrum, a direct electronic bandgap energy of 3.6(1) eV and the signature of an indirect electronic transition around 0.5 eV are obtained. These values provide necessary experimental feedback to previous conflicting electronic band structure calculations predicting either a topologically trivial or non-trivial insulating ground state in YBiO$_3$.
\end{abstract}


\maketitle


\section{Introduction}
During the past ten years, heteroepitaxial YBiO$_3$ thin films were investigated mainly as a buffer layer for epitaxial YBa$_2$Cu$_3$O$_{7-\delta}$ high-$T_\mathrm{c}$ superconductors.\cite{Li2007,Zhao2007,Pollefeyt2013,Meiqiong2009} Recently, first-principles electronic band structure calculations proposed YBiO$_3$ as a novel oxide topological insulator:\cite{Jin2013} Based on an assumed, undistorted perovskite structure with space group $Pm\bar{3}m$ and lattice constant $a$~=~5.428~\r{A}, a topologically insulating phase was predicted due to a band inversion between an $s$-like and a $j_\mathrm{eff}$~=~1/2 band at the $R$ point that is driven by the spin-orbit coupling of the Bi $6p$ states. YBiO$_3$ then has a topological trivial indirect bandgap of $\approx$~0.18~eV and a non-trivial direct bandgap of 0.33~eV at the $R$ point. The size of the direct bandgap and high bulk resistivity were expected to allow for surface-dominated transport at room-temperature. However, a subsequent theoretical study showed that the assumed $Pm\bar{3}m$ crystal structure is unstable.\cite{Trimarchi2014} Instead, a distorted $Pnma$ perovskite structure is predicted to be most stable and YBiO$_3$ is predicted as a topologically trivial band insulator with larger direct and indirect electronic bandgaps.

Experimentally, solid solutions of Y$_x$Bi$_{1-x}$O$_{1.5}$ with $0.1<x<0.5$ are stable at room-temperature and crystallize in a face-centered cubic, defective fluorite-type structure with space group $Fm\bar{3}m$.\cite{Battle1983,Battle1986,Kale1999,Zhang2010a} However, there exist no band structure calculations for the $Fm\bar{3}m$ space group. At $x$~=~0.5, Y$_x$Bi$_{1-x}$O$_{1.5}$ (YBiO$_3$) has a cubic lattice constant $a$~=~5.4188~\r{A},\cite{Zhang2010a} and its pseudo-cubic primitive cell with $a/\sqrt{2}$ is closely matched to LaAlO$_3$, SrTiO$_3$, (LaAlO$_3$)$_0.3$(Sr$_2$TaAlO$_6$)$_0.7$ (LSAT) and also YBa$_2$Cu$_3$O$_{7-\delta}$. Heteroepitaxial YBiO$_3$(100) thin films have hence been prepared as buffer layers on LaAlO$_3$(001), SrTiO$_3$(001) and LSAT(001) single-crystalline substrates by chemical solution\cite{Li2007,Zhao2007,Pollefeyt2013} and pulsed laser deposition.\cite{Meiqiong2009,dePutter2014} Their surface-morphology is characterized by crystalline grains of around 50~nm and root-mean-squared (RMS) surface roughnesses between 2.6 to 1.8~nm at a film thickness of 140 and 40~nm, respectively.\cite{Zhao2007,Pollefeyt2013} 
Furthermore, pelletized YBiO$_3$ powders showed a high 0.5 M$\Omega$~m bulk resistivity at room-temperature and paramagnetic behavior\cite{Zhao2007}.

In order to provide some necessary experimental feedback to the conflicting theoretical band structure calculations,\cite{Jin2013,Trimarchi2014} we report here an estimate of the direct electronic bandgap of two relaxed-textured YBiO$_3$(111) films deposited using pulsed laser deposition with different thickness on $a$-Al$_2$O$_3$ single crystals. Structural analyses confirm their crystallization in a defective fluorite-type structure with $Fm\bar{3}m$ space group. The dielectric function between 0.03~eV and 4.5~eV is determined by simultaneous modeling of spectroscopic ellipsometry and optical transmission data. Contrary to the theoretical band structure calculations, we find evidence for a direct electronic bandgap of 3.6(1)~eV and a possible signature of an indirect electronic transition around 0.5~eV in YBiO$_3$.

\section{Experimental details}
The YBiO$_3$ thin films were grown by pulsed laser deposition (PLD) on 10~$\times$~10~mm$^2$ $a$-plane Al$_2$O$_3$(11-20) single crystals. PLD was performed with a 248~nm KrF excimer laser at a laser fluence of 2~J~cm$^{-2}$. The polycrystalline source target was prepared by conventional solid-state synthesis of Bi$_{2}$O$_3$ and Y$_2$O$_3$ powders in a 1:1 molar ratio. The starting materials were homogenized, pressed and sintered in air for 24~h at 880~$^\circ$C. After an intermediate regrinding, a second and third sinter step was performed for 12 h in an oxygen atmosphere at 800~$^\circ$C and 1,000~$^\circ$C, respectively.

The deposition process involved a 300 laser pulses nucleation layer grown at a 1~Hz pulse frequency, followed by the deposition of 60,000 and 2,000 pulses at 5~Hz for the thick and thin film, respectively. The films were grown at a growth temperature of approximately 650~$^\circ$C and an oxygen partial pressure of 0.05 mbar to obtain optimal Y:Bi stoichiometry and film crystallinity. After the deposition, the samples were annealed \emph{in situ} at an oxygen partial pressure of 800~mbar. The film thicknesses were determined as 1660~nm and 57~nm from optical modeling (see text below). Note, that reliable resistivity data was not obtained due to a very large bulk resistance of the 1660~nm film of around 700~M$\Omega$ at room-temperature which exceeds our instrument's measuring range.

X-ray diffraction (XRD) structural analyses were performed with a Panalytical X’Pert PRO Materials Research Diffractometer with parabolic mirror and PIXcel$^{3D}$ detector and Cu K$_\mathrm{\alpha}$ radiation. The surface morphology was investigated with a Park Systems XE-150 atomic force microscope (AFM) in dynamic non-contact mode. Topographic images were post-processed with the Gwyddion software.\cite{Necas2012} The chemical composition was measured by energy-dispersive X-ray spectroscopy (EDX) using an FEI Novalab 200 scanning electron microscope (SEM) equipped with an Ametek EDAX detector. The thin film dielectric function and layer thickness were determined via standard variable angle spectroscopic ellipsometry (VASE, IR-VASE by J.A. Woolam, Inc.) in the spectral range from 0.03~eV to 4.50~eV and UV/VIS transmission spectroscopy (Perkin Elmer Lambda 19) between 0.62~eV and 6.20~eV.

\section{Results and discussion}
\subsection{Structural properties}
\begin{figure}
\includegraphics[width=\columnwidth]{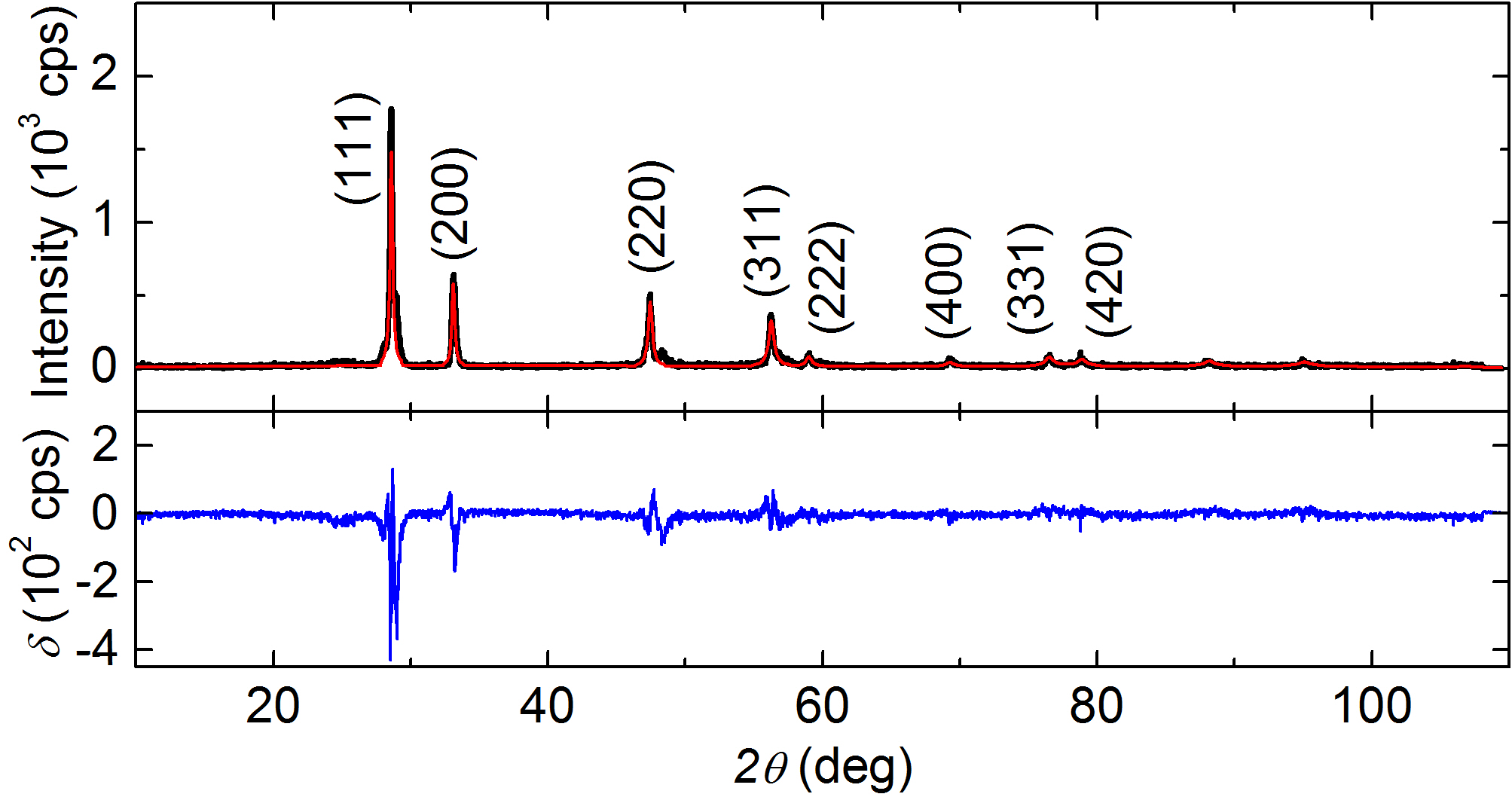}
\caption{\label{fig:figure1}(Color online) Top: XRD $2\theta$-$\omega$-scan (black) of the polycrystalline YBiO$_3$ source target. The Rietveld method\cite{Rietveld1969} is applied using the defective fluorite-type $Fm\bar{3}m$ structural model\cite{Zhang2010a} with $a$~=~5.4279(4)~\r{A}. Bottom: Difference $\delta$ between observed and calculated intensities.}
\end{figure}

Figure \ref{fig:figure1} shows an XRD $2\theta$-$\omega$-scan of the polycrystalline YBiO$_3$ source target. The pattern is fit successfully by the Rietveld method\cite{Rietveld1969} using the defective fluorite-type $Fm\bar{3}m$ structural model by Zhang \emph{et al.}\cite{Zhang2010a} and a cubic lattice constant $a$ = 5.4279(4) \r{A}. There exist minor additional peaks that very likely relate to the unreacted starting materials, and possibly elemental Bi and Bi$_2$O$_{3-x}$.\cite{Pollefeyt2013} We note, that the assumed $Pm\bar{3}m$ and $Pnma$ models fail to match with all of the observed YBiO$_3$ reflexes, see Supplementary Material. EDX yields a Y:Bi ratio of 0.98:1 which suggests that good stoichiometric transfer is feasible in spite of the incomplete target phase purity.

\begin{figure}
\includegraphics[width=\columnwidth]{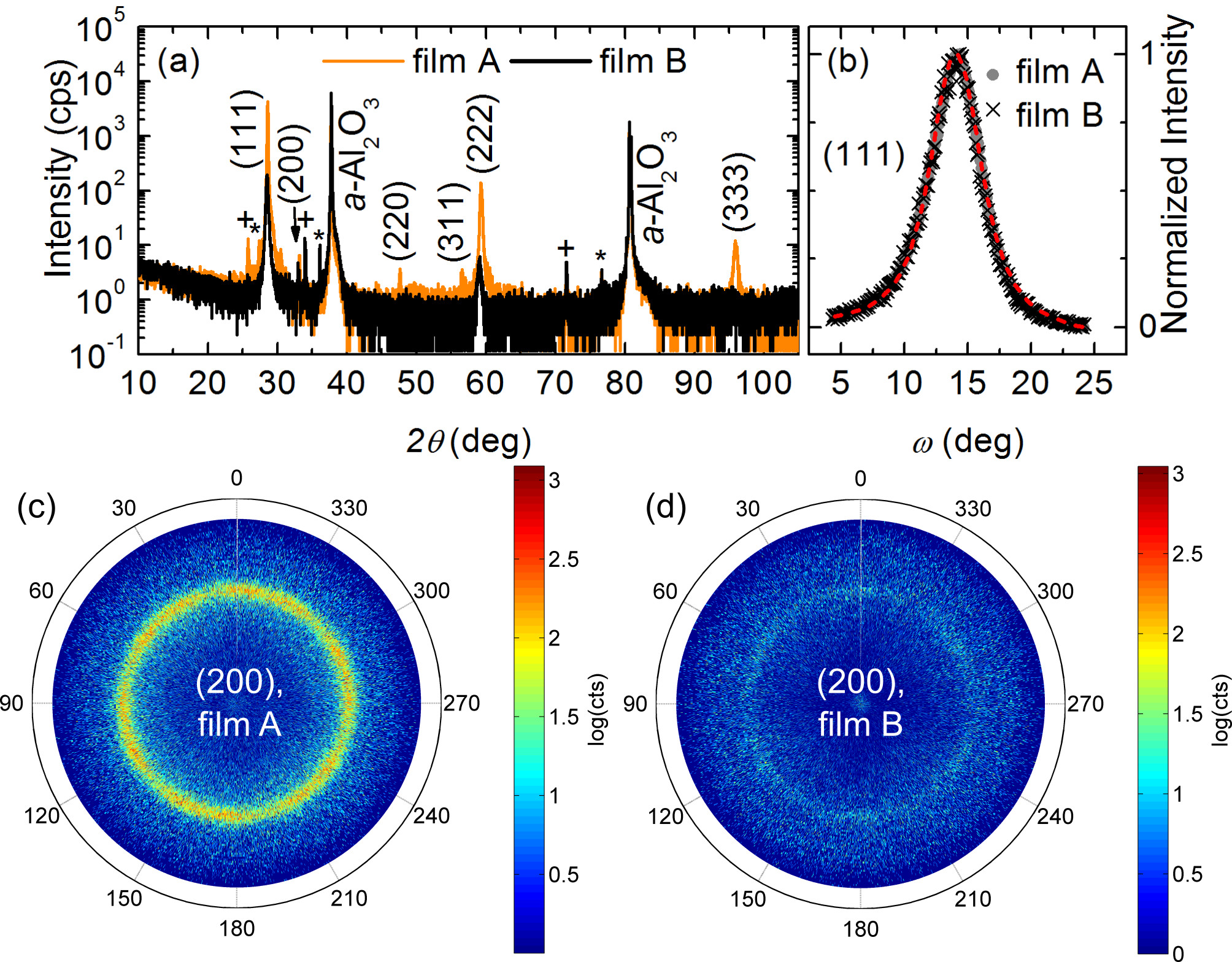}
\caption{\label{fig:figure2}(Color online) XRD performed on 1660~nm and 57~nm YBiO$_3$ films A and B deposited on $a$-Al$_2$O$_3$: (a) $2\theta$-$\omega$-scans confirm an excellent (111) preferential out-of-plane orientation. Cu K$_{\mathrm{\beta}}$ and W lines are denoted by $+$ and $*$. (b) Normalized $\omega$-scans of the YBiO$_3$ (111) symmetric reflex with identical half-widths of 4.94(4)$^\circ$ obtained by a Gauss fit (red dashed line). (c,d) $\phi$-$\chi$ pole figures of the YBiO$_3$ (200) reflex indicate a nearly random distribution of the crystallite in-plane orientation in both films.}
\end{figure}

Figure \ref{fig:figure2}(a) shows the $2\theta$-$\omega$ patterns of two YBiO$_3$ films with a thickness $d$ of 1660 nm (film A) and 57 nm (film B), respectively, deposited on $a$-Al$_2$O$_3$ single crystals. The patterns are indexed according to the experimental $Fm\bar{3}m$ unit cell\cite{Battle1983,Battle1986,Kale1999,Zhang2010a} and indicate very good (111) preferential out-of-plane orientation. For film A, the pattern shows additional minor peaks ($<10$~cps) related to other allowed YBiO$_3$ reflexes. By extrapolation of the $\theta$ values of the (111), (222) and, if possible, the (333) reflexes to $\theta=0^{\circ}$,\cite{Nelson1945} very similar out-of-plane lattice constants $a_\mathrm{out}$ of 5.3907(2)~\r{A} and 5.42(4)~\r{A} are determined for films A and B, respectively. These values deviate from the literature value of 5.4188 \r{A} \cite{Zhang2010a} and from the Rietveld-refined value of the polycrystalline target by at most 0.5 and 0.7~\%, respectively.

For both films, the XRD $\omega$-scans (rocking scans) of the YBiO$_3$ (111) reflex, see Fig. \ref{fig:figure2}(b), exhibit identical half-widths of about 5$^\circ$ and thus suggest a comparable degree of crystallite mosaicity.

A possible in-plane epitaxial relationship with $a$-Al$_2$O$_3$ was investigated by $\phi$-$\chi$ pole figures around the asymmetric YBiO$_3$ (200) reflex, see Fig. \ref{fig:figure2}(c,d). For both films A (c) and B (d), a distinct circular intensity maximum is evidence for a random distribution of the in-plane crystallite orientation. The lack of a defined in-plane epitaxial relationship is explained by the large lattice mismatch between YBiO$_3$(111) and $a$-Al$_2$O$_3$ of about $\pm$40 \%.\footnote{Assuming a cubic YBiO$_3$(111) epilayer with $a_\mathrm{(111)}=\sqrt{2}\cdot 5.42$ \r{A}, the lattice mismatch with the 13.00~$\times$~5.50~\r{A}$^2$ $a$-plane of Al$_2$O$_3$ will be +42~\% and -40~\%, respectively.}

The XRD data indicate relaxed, i.e., nominally unstrained growth of YBiO$_3$(111) on $a$-Al$_2$O$_3$. Thus, the observed preferential out-of-plane orientation suggests, that the YBiO$_3$ (111) surface has the lowest formation energy. Because EDX analyses on both films yield nearly ideal Y:Bi ratios of 1.02:1, the slight deviations of the out-of-plane lattice constant from the literature \cite{Zhang2010a} and the polycrystalline target value are likely caused by internal strain due to, e.g., oxygen vacancies.

\subsection{Optical properties}
The spectra of the dielectric function were determined from ellipsometry and optical transmission data of the relaxed grown YBiO$_3$ films A and B. The anisotropic transfer matrix model consists of layers for the YBiO$_3$ film and the surface roughness using a Bruggeman effective medium approximation mixing the dielectric functions of YBiO$_3$ and air in a 1:1 ratio.\cite{Jellison1994} The dielectric functions of the films are modeled using a Kramers-Kronig consistent numerical B-spline model.\cite{Johs2008} We note, that film A was deposited on a double-sided polished substrate to avoid diffusive scattering in the optical transmission spectroscopy. The substrate backside was subsequently roughened mechanically in order to minimize back surface reflections in ellipsometry. The substrate of film B was single-sided polished. Its wavelength-dependent scattering by the backside roughness in the transmission measurement is corrected for by a reference transmission measurement of a bare double-sided polished substrate with same thickness. Finally, the absorption coefficient $\alpha$ is calculated from the numerical B-spline dielectric function.

\begin{figure*}
	\centering
		\includegraphics[width=\textwidth]{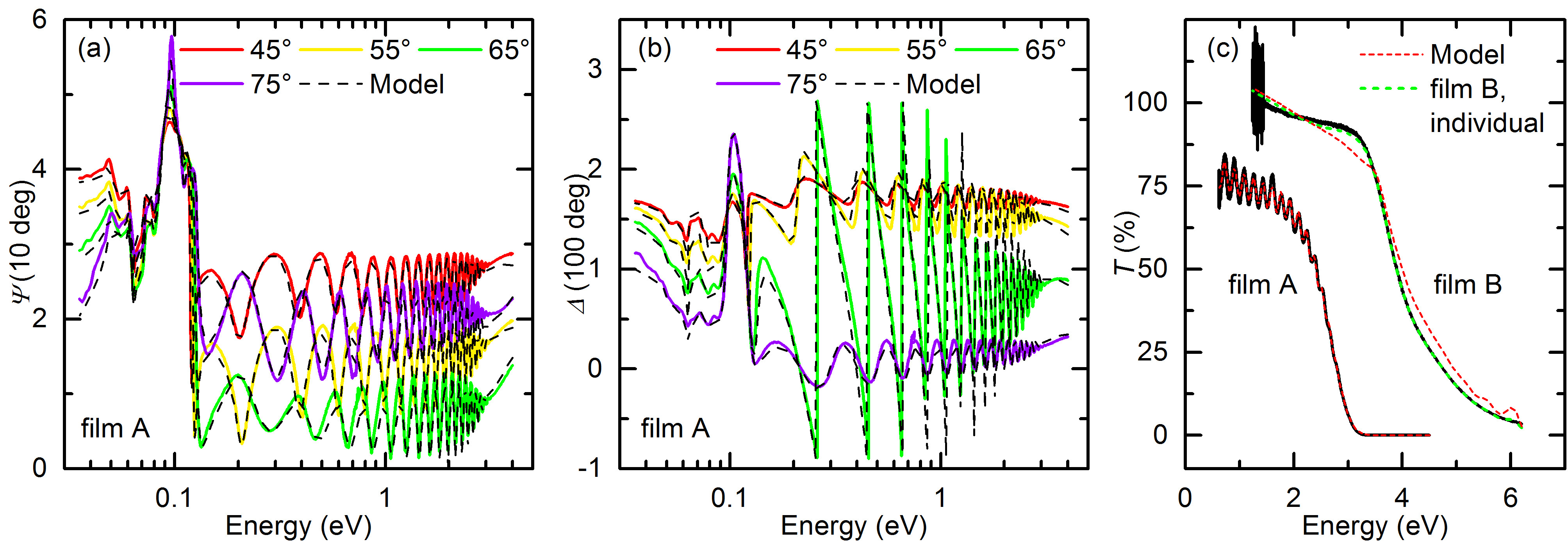}
	\caption{(Color online) Measured and model-calculated spectra of the ellipsometric parameters $\mathit{\Psi}$ (a) and $\mathit{\Delta}$ (b) of the 1660~nm thick YBiO$_3$ film A on $a$-Al$_2$O$_3$ measured at angles of incidence as indicated. For $\mathit{\Psi}$ and $\mathit{\Delta}$ of the 57 nm film B, we refer to the Supplementary Material. (c) Optical transmission spectra $T$ of films A and B (black) are shown together with the corresponding simultaneous model spectra (red). For comparison, the individual model spectrum of film B (green) is shown as well.}
	\label{fig:YBO_DF}
\end{figure*}

Figure \ref{fig:YBO_DF}(a-c) shows the results of the simultaneous modeling of the reflection ellipsometry ($\mathit{\Psi}$, $\mathit{\Delta}$) and optical transmission ($T$) data of the YBiO$_3$ films A and B. The numerical B-spline model reproduces the experimental data in the entire measured spectral range (a,b). Around 200~meV, substrate backside reflections disturb the ellipsometry spectra. These occur only in this narrow spectral range of large wavelengths where film and substrate are sufficiently transparent. At lower photon energies, the $a$-Al$_2$O$_3$ Restrahlen band prevents this. The modeled $T$ data of film B, Fig. \ref{fig:YBO_DF}(c), suffers from a large error as it was measured through a rough backside; in this case, the ellipsometry data is more reliable. The spectra of film A show obvious layer thickness oscillations. For films A and B, thicknesses of $d$~=~1660(30)~nm and $d$~=~57(2)~nm are obtained from the fits. The modeled effective surface roughnesses are approximately 25~nm and 3~nm, respectively. These values agree well with surface roughness values obtained via AFM, see Supplementary Material.

\begin{figure}
	\centering
		\includegraphics[width=1\columnwidth]{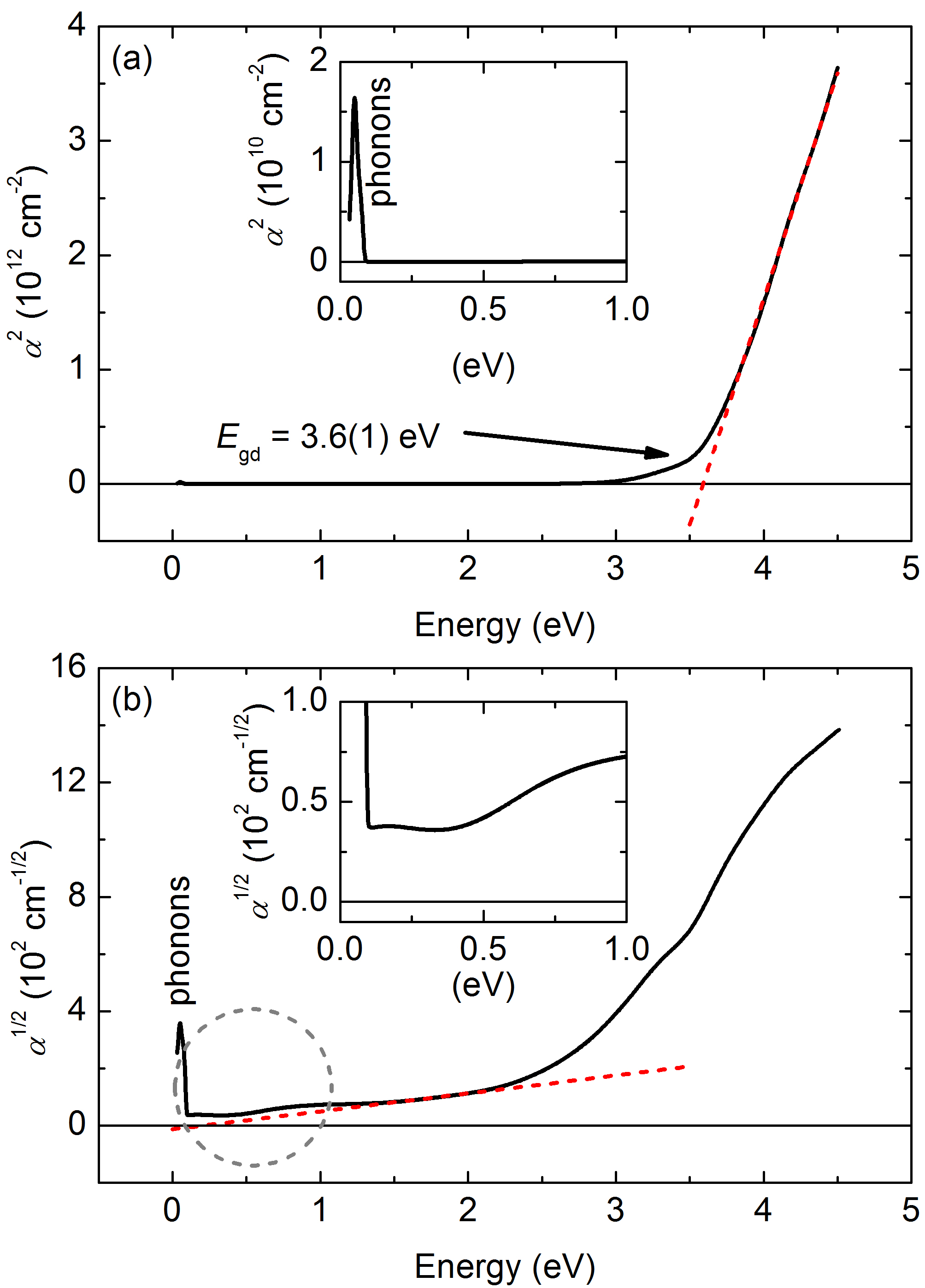}
	\caption{(Color online) Calculated absorption spectra plotted versus photon energy as $\alpha^{2}$ (a) and $\alpha^{1/2}$ (b). (a) Linear regression of $\alpha^{2}$ between 3.75~eV and 4.50~eV and extrapolation to $\alpha^{2}$~=~0 yields an estimate of the direct bandgap of $E_\mathrm{gd}$~=~3.6(1)~eV. (b) The $\alpha^{1/2}$ spectrum exhibits a phonon resonance at about 0.05~eV and possible indirect bandgap absorption around 0.5~eV. The insets in (a) and (b) show a zoom-in of the 0~eV to 1~eV spectral range.}
	\label{fig:E4048_alphafits}
\end{figure}

From the numerical model dielectric function the absorption coefficient $\alpha$ is calculated and plotted in Fig.~\ref{fig:E4048_alphafits} as $\alpha(E)^2$ and $\alpha(E)^{1/2}$, respectively. Based on the present data, we can exclude indirect and direct bandgaps at $\approx$~0.183~eV and 0.33~eV, as predicted by the band structure calculations of Jin \emph{et al.}\cite{Jin2013} Instead, strong absorption is evident above 3~eV together with a continuous onset at about 0.5~eV. At about 0.05~eV, a phonon resonance is evident. The strong absorption peak is interpreted in analogy to Trimarchi \emph{et al.} \cite{Trimarchi2014} as the allowed direct bandgap transition between the Bi $6s$ and $6p$ bands at the $R$ point. For an estimate of the direct bandgap energy $E_\mathrm{gd}$, linear regression of the $\alpha^2$ spectrum, Fig.~\ref{fig:E4048_alphafits}(a), between 3.75~eV and 4.50~eV is performed.\cite{Fujiwara2007} The extrapolation of the straight line fit to 0 yields $E_\mathrm{gd}$~=~3.6(1)~eV. Furthermore, we tentatively associate the slight absorption starting at around 0.5~eV, visible in Fig. \ref{fig:E4048_alphafits}(b), with the allowed indirect bandgap transition from the valence band maximum at the $\Gamma$ point to the conduction band minimum at the $R$ point.\cite{Trimarchi2014} To cross-check the reliability of our model, we have also modeled films A and B individually, as shown in the Supplementary Material: While the model of the thin film B yields the same direct bandgap, cf. Fig.~\ref{fig:YBO_DF}(c), but is insensitive to low-energy absorption, modeling the thick film A is required to reveal a possible Urbach tail between 2.25~eV and 3.25~eV.

The obvious discrepancy between the predicted \cite{Jin2013,Trimarchi2014} and experimentally observed direct bandgap energy in YBiO$_3$ is not fully understood. Density functional theory models of a material's band structure typically underestimate the real bandgap by up to 40~\%.\cite{Perdew2009} However, the obtained bandgaps of the nominally unstrained YBiO$_3$ films seem reasonable as they linearly interpolate between the end members of the Y$_\mathrm{x}$Bi$_\mathrm{1-x}$O$_{1.5}$ solid solutions: While as grown thin films of cubic $\delta$-Bi$_2$O$_3$ have an indirect optical gap of 1.73~eV,\cite{Fan2006} pulsed laser-deposited cubic Y$_2$O$_3$ thin films exhibit a direct bandgap of 5.6~eV.\cite{Zhang1998} In this light, bowing effects in the band gap appear to be small.

\section{Summary}
In summary, we have prepared heteroepitaxial YBiO$_3$ films on $a$-Al$_2$O$_3$ single crystals by pulsed laser deposition. Rietveld refinement of the polycrystalline source target confirms that the defective fluorite-type structure with space group $Fm\bar{3}m$\cite{Battle1983,Battle1986,Kale1999,Zhang2010a} is the correct structural model for YBiO$_3$. At film thicknesses of both 1660~nm and 57~nm, a (111) preferential out-of-plane orientation is obtained. X-ray diffraction analyses also confirm a relaxed and nominally unstrained film growth. In particular, there exists a random distribution of the in-plane crystallite orientation. The dielectric function is determined in the 0.03~eV to 4.50~eV spectral range by simultaneous modeling of spectroscopic ellipsometry and optical transmission data of both film samples. From the calculated absorption coefficient a direct electronic bandgap of 3.6(1)~eV is obtained. We furthermore find a possible signature of another, indirect electronic transition around 0.5~eV. Together these values provide necessary experimental feedback to electronic band structure calculations that have proposed either a topologically trivial\cite{Trimarchi2014} or a non-trivial\cite{Jin2013} insulating ground state.

\section{Supplementary Material}
See Supplementary Material for further experimental details on Rietveld refinement, AFM and SEM images, and the simultaneous numerical B-spline model dielectric function.

\begin{acknowledgments}
We thank the Deutsche Forschungsgemeinschaft (DFG) for financial support within the collaborative research project SFB 762 "Functionality of oxide interfaces". S.R. is grateful to the Leipzig Graduate School of Natural Sciences BuildMoNa.
\end{acknowledgments}

\bibliography{PaperYBO}

\clearpage
\widetext

\section*{{\large Supplemental Material: Fundamental absorption edges in heteroepitaxial YBiO\textsubscript{3} thin films}}

\fontsize{12}{24}\selectfont

\setcounter{section}{0}
\setcounter{equation}{0}
\setcounter{figure}{0}
\setcounter{table}{0}
\setcounter{page}{1}
\makeatletter

\renewcommand{\theequation}{S\arabic{equation}}
\renewcommand{\thefigure}{S\arabic{figure}}
\renewcommand{\bibnumfmt}[1]{[S#1]}
\renewcommand{\citenumfont}[1]{S#1}

\section{Rietveld analysis of the polycrystalline YBiO$_3$ PLD source target}
\begin{figure}[b]
\includegraphics[width=\columnwidth]{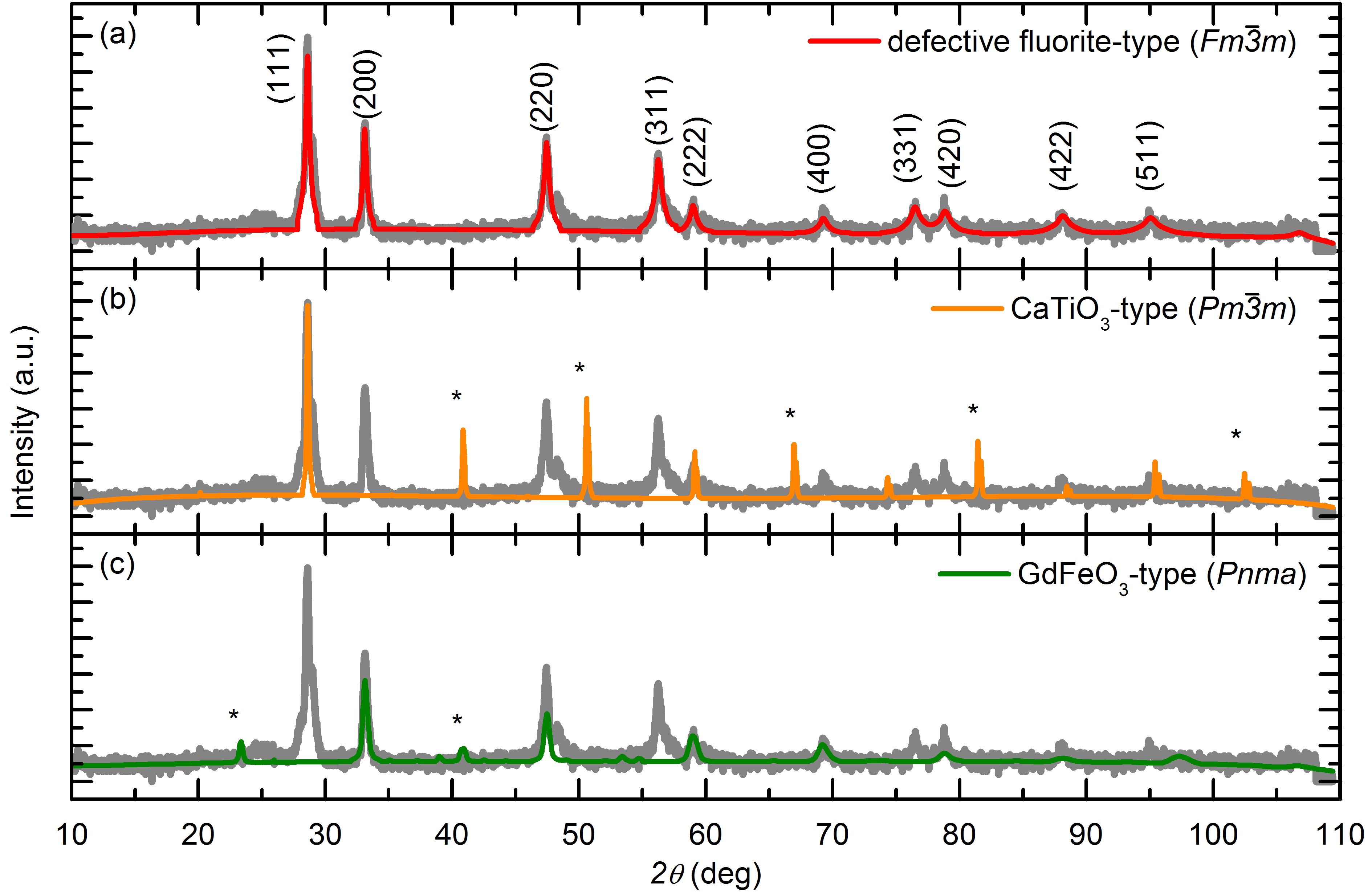}
\caption{\label{fig:Rietveld}(Color online) The XRD $2\theta$-$\omega$ pattern (gray) of the polycrystalline YBiO$_3$ PLD source target was Rietveld refined using (a) the defective fluorite-type $Fm\bar{3}m$, \cite{Zhang2010a} (b) the CaTiO$_3$-type (ICSD 31865, $a$ = 4.405 \r{A} \cite{Trimarchi2014}), and (c) the GdFeO$_3$-type (ICSD number 16688) structural models. Bragg reflexes not observed in the pattern are denoted by a $*$. The refined lattice parameters are (a) $a$~=~5.4279(4)~\r{A}, (b) $a$ = 4.420(1) \r{A}, and (c) $a$~=~5.461(5)~\r{A}, $b$~=~7.711(8)~\r{A}, $c$~=~5.415(5)~\r{A}. Goodness of fits $\chi^{2}$: 3.98 (a), 20.99 (b), 6.91 (c).}
\end{figure}

Figure \ref{fig:Rietveld} shows the XRD $2\theta$-$\omega$-scan of the polycrystalline YBiO$_3$ PLD source target. The pattern is fit by the Rietveld method\cite{Rietveld1969} using the defective fluorite-type $Fm\bar{3}m$ structure\cite{Zhang2010a} with cubic lattice constant $a$ = 5.4279(4) \r{A}, and the assumed CaTiO$_3$-type $Pm\bar{3}m$ and GdFeO$_3$ $Pnma$ structural models.\cite{Trimarchi2014} The assumed structures fail to simultaneously match with all of the observed YBiO$_3$ Bragg reflexes, such as the YBiO$_3$ (111) and (200) planes measured also in thin film samples. Only the $Fm\bar{3}m$ model explains all experimentally observed Bragg reflexes.

\section{Surface morphology}
\begin{figure}
\includegraphics[width=1.00\textwidth]{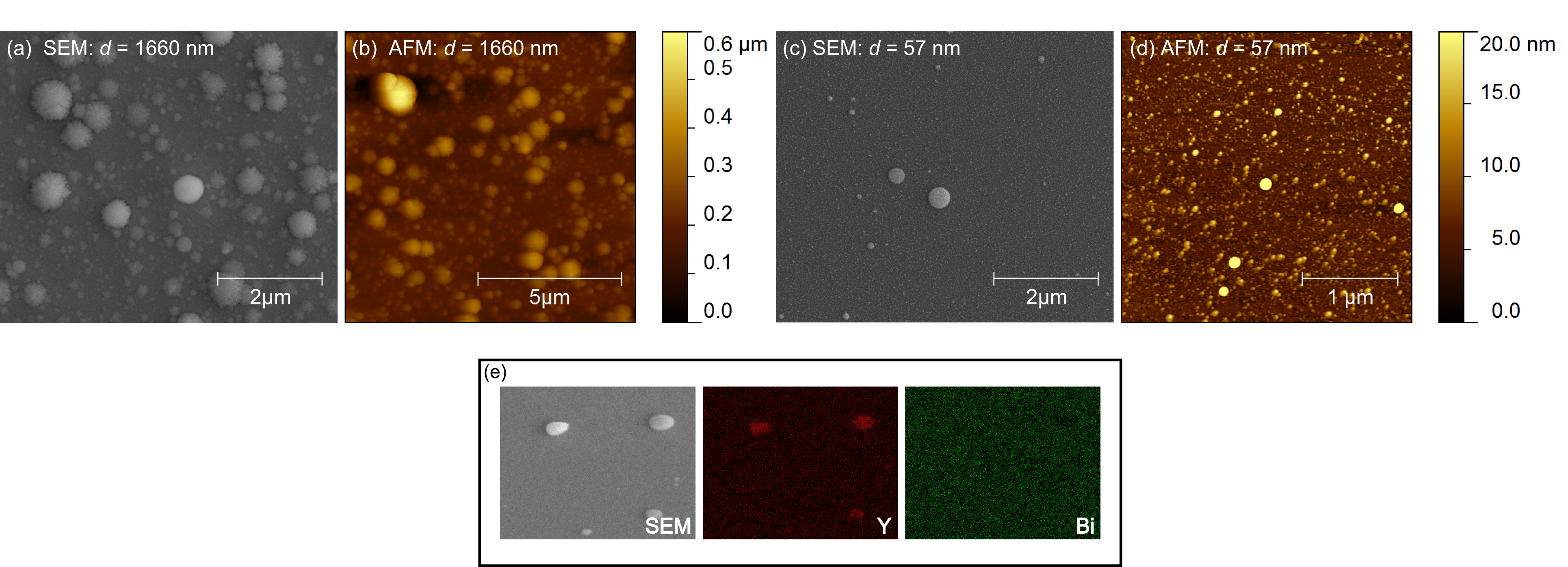}
\caption{\label{fig:AFM}(Color online) SEM and non-contact AFM topographic images of the 1660~nm (a,b) and 57~nm (c,d) YBiO$_3$ films A and B. (e) SEM and EDX Y and Bi mapping of typical droplets.}
\end{figure}

The surface morphology was investigated with a Park Systems XE-150 atomic force microscope (AFM) in dynamic non-contact mode and an FEI Novalab 200 scanning electron microscope (SEM). Topographic AFM images were post-processed with the Gwyddion software.\cite{Necas2012} Figure \ref{fig:AFM}(a-d) shows topographic AFM and SEM images of the 1660 nm (a,b) and 57~nm (c,d) YBiO$_3$ films A and B. The surface of film A is characterized by separate, very large crystalline grains with lateral dimensions around 700~nm and droplets. From the AFM image, a root-mean-squared (RMS) surface roughness of 55.8~nm is obtained. Droplet formation is also evident on the surface of film B (b) with an RMS surface roughness of 2.68~nm, and crystalline grain dimensions between 30~nm and 80~nm. The surface roughness and grain size are compatible with other YBiO$_3$ films.\citep{Li2007,Zhao2007,Pollefeyt2013} EDX analysis revealed that the droplets are mainly composed of yttrium. We argue, that the observed surface morphology can be explained by droplets 
and a mechanically soft target.

\section{Dielectric function}
\begin{figure}
\includegraphics[width=1.00\textwidth]{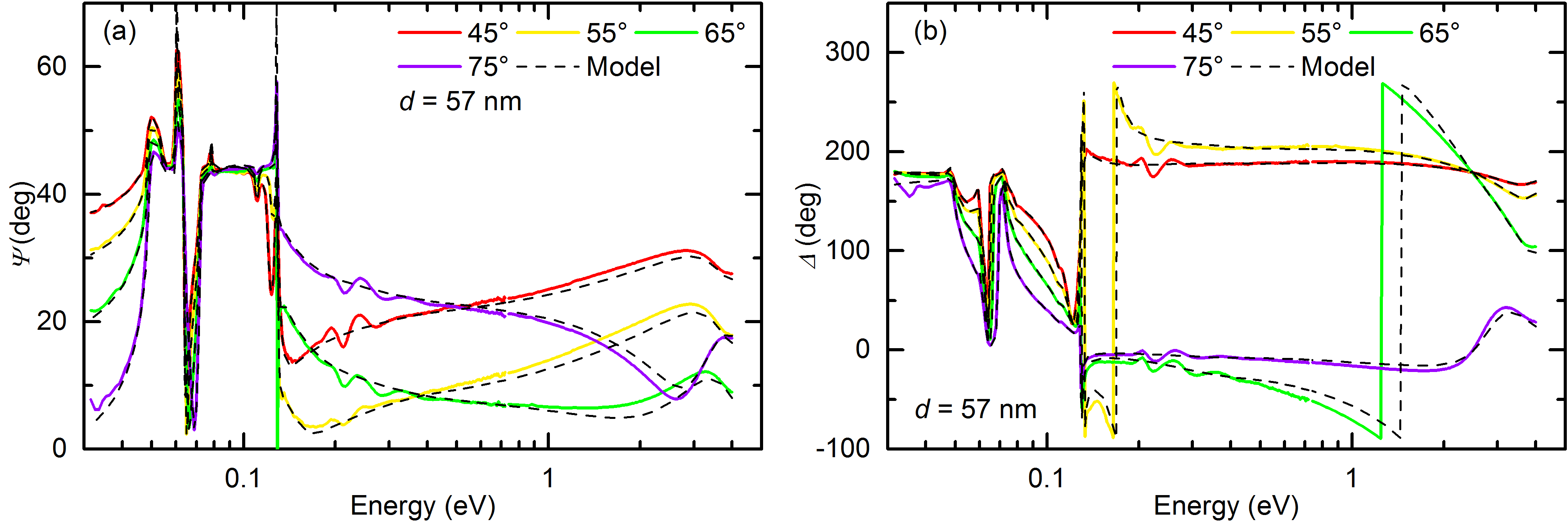}
\caption{\label{fig:DF}(Color online) Measured and model-calculated spectra of the ellipsometric parameters $\mathit{\Psi}$ (a) and $\mathit{\Delta}$ (b) of 57~nm thick YBiO$_3$ (film B) measured at angles of incidence as indicated.}
\end{figure}

The dielectric function of YBiO$_3$ was determined by means of standard variable angle spectroscopic reflection ellipsometry in the 0.03~eV to 4.5 ~eV spectral range applied to the two 1660~nm and 57~nm YBiO$_3$ film samples A and B. For optically isotropic YBiO$_3$, the measured quantities are the ellipsometric parameters $\mathit{\Psi}$ and $\mathit{\Delta}$, which are defined by the ratio $\rho=r_\mathrm{p}/r_\mathrm{s}=\tan\mathit{\Psi}\exp (i\mathit{\Delta})$ of the p- and s-polarized complex reflection coefficients $r_\mathrm{p}$ and $r_\mathrm{s}$. A transfer matrix model is simultaneously adapted to the ellipsometry and optical transmission data to obtain a numerical B-spline model dielectric function of both YBiO$_3$ films.\cite{Johs2008} The measured and modeled $\mathit{\Psi}$ and $\mathit{\Delta}$ spectra of the 57 nm YBiO$_3$ film are shown in Figure~\ref{fig:DF} (for film A see Fig.~3 of the main the text).

\begin{figure}
\includegraphics[width=1.00\textwidth]{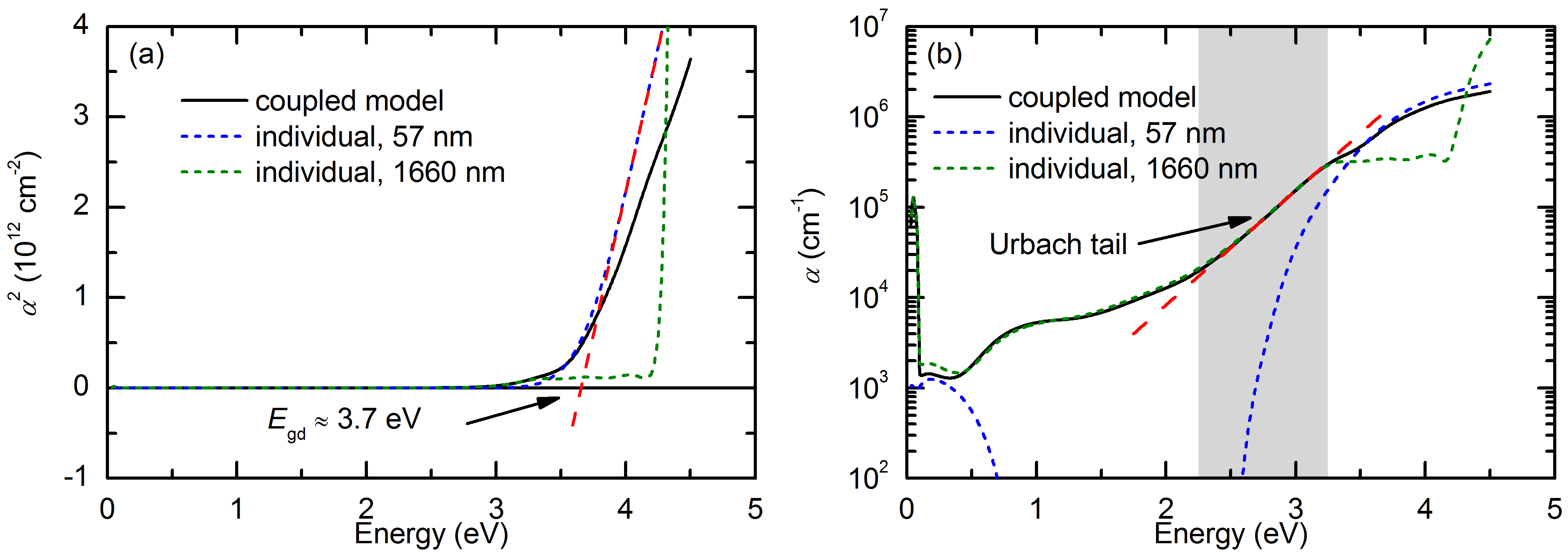}
\caption{\label{fig:alphas}(Color online) Absorption spectra calculated from simultaneous modeling (black) and individual modeling of YBiO$_3$ films A (green) and B (blue) deposited on $a$-Al$_2$O$_3$. The data are plotted as $\alpha^{2}$ (a) and $\alpha$ (b, log-lin scale) versus incident photon energy.}
\end{figure}

Figure \ref{fig:alphas} displays the calculated absorption coefficient $\alpha$ obtained from the simultaneous modeling (cf. Fig. \ref{fig:DF}) and from individual models of YBiO$_3$ films A and B, respectively. From the simultaneous modeling a direct bandgap of 3.6(1) eV was estimated. A similar result of $\approx$~3.7~eV is obtained from the individual model of film B, which is however insensitive to absorption at lower energies. In contrast, film A is required to correctly reproduce absorption below about 3.3 eV, but is insensitive to absorption at higher energies. The model of film A furthermore exhibits an additional Urbach tail between 2.25 eV and 3.25 eV, as indicated by a straight line in the semi-log plot of $\alpha$,\cite{Studenyak2014} Fig. \ref{fig:alphas}(b). In summary, the simultaneous modeling of films A and B is required to reveal the direct and possible indirect bandgap in YBiO$_3$.

\bibliography{PaperYBO}

\end{document}